\documentclass[preprint, 12pt]{elsarticle}

\usepackage{graphicx}
\usepackage{amssymb}
\usepackage{xcolor}

\biboptions{sort&compress}

\journal{Journal of Computational Materials Science}

\begin{document}

\begin{frontmatter}


\title{High-velocity impact of hybrid CBN nanotube}



\author{Enzo Armani}

\address{Center of Natural Human Science Federal University of ABC – UFABC, Santo Andre Brazil}

\author{Pedro A. S. Autreto} 

\address{Center of Natural Human Science Federal University of ABC – UFABC, Santo Andre Brazil}

\begin{abstract}

Nanomaterials under extreme conditions can behave in a completely different manner. High-velocity impact, for example, can produce nanoribbons without any chemical approach via carbon or boron nitride nanotubes unzipping. Although hybrid nanostructures have been used to create stronger structures, few studies on these materials under extreme conditions have been employed. In this work, we study an experimentally synthesized hybrid doubled walled CBN nanotube (boron nitride and carbon nanotubes concentrically assembled) under high-velocity impact. Our results show that the combination of elastic and brittle materials can produce different structures, such as nanoribbons and boron nitride atomic chains. These results can have a significant impact on the production of new nanostructures. 
\end{abstract}


\end{frontmatter}


\section{Introduction}

\label{Intro}

Carbon-based nanomaterials have been in decades one of the highlights on materials science studies \cite{geim2010rise, geim2009graphene}. It is highly due to the carbon atom hybridization types ($sp$, $sp^2$ and $sp^3$) which allows nanostructures in all dimensionalities as for example fullerenes (0D) \cite{geim2010rise, curl1991fullerenes}, carbon nanotubes (CNT)(1D) \cite{iijima1993single}, graphene (2D) \cite{novoselov2004electric, novoselov2005two}, and nanoscrolls (3D) \cite{braga2004structure}. Those structures can perform a plethora of different thermal, electronic and mechanical properties and have been employed in many different technological applications among petroleum industry \cite{vinod2014low}, aerospace \cite{paipetis2012carbon} and biological systems \cite{vardharajula2012functionalized} in many different conditions.

Particularly interesting are the nanostructures under extreme conditions that have been extensively studied in recent times \cite{ozden2014unzipping, de2016carbon}.  The combination of experiment and molecular dynamics simulations of the high-velocity impact of nanotubes on solid targets has presented that this method can provide stress concentration in a specific region of this slender structure \cite{ozden2014unzipping}. This unexpected result creates an innovative way to obtain carbon nanoribbons from CNT unzipping.

Therefore, such mechanical response of CNT under high-velocity impact collision stimulated studies involving other materials \cite{machado2016structural, de2016carbon}. Hexagon boron nitride (hBN), which also shares the same hexagonal honeycomb lattice \cite{perim2013dynamical} and can form tubes (BNNT), behaves remarkably different under high strain deformation rates. Multi-walled BNNT unzipping occurs via an almost linear fracture on the top side (the opposite side of the collision) because of the hardness of the inner core \cite{machado2016structural}. In this way, this process can open only one tube at a time. On the other hand, multi-walled CNT (MWCNT) unzips completely (both walls) by C-C breaking bonds on the CNT edges through the stress concentration. This is an effect of faster stress propagation than in B-N and more flexibility of the inner wall   \cite{machado2016structural, perim2013dynamical, dos2012unzipping}.

\begin{figure}[b]
   \centering
    \includegraphics[width=15.0cm] {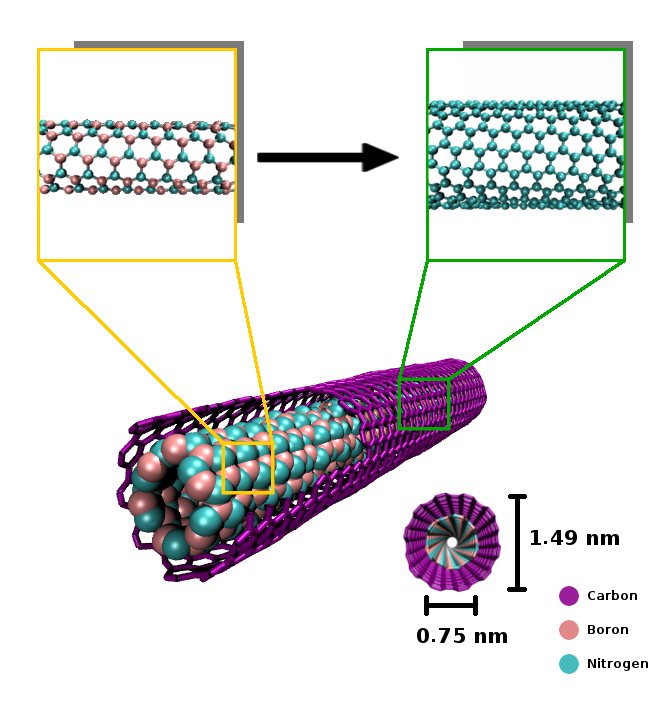}
    \caption{Hybrid nanotube (HNT): a concentrically composition of single-wall boron nitride nanotube (in red and cyan, respectively), inside of a carbon nanotube (in purple).}
    \label{fig1}
\end{figure}

A good question is how a structure composed by CNT and BNNT concentrically disposed (hybrid nanotube - HNT) will behave under similar conditions (high-velocity impact) given that CNT and BNNT present distinct fractures types. HNT has been already successfully synthesized in 2013 using ammonia borane complexes (ABC) \cite{nakanishi2013thin} in a CNT template. This HNT consists of a BNNT, with a diameter of 0.75nm, inside a 1.49nm diameter CNT (with chiralities of (6,5) and (12,10), respectively), forming a multi-wall nanotube (Figure \ref{fig1}).  In this work, we study the fracture of HNT under extreme conditions (high-velocity impact on a solid surface) using fully atomistic reactive molecular dynamics simulations. 

\section{Methodology}

To study the high-velocity impact of the HNT under different conditions, we carried out fully atomistic reactive molecular dynamics (MD) simulations, using the Reactive Force Field (ReaxFF)\cite{van2001reaxff}, a force-field that allows breaking bonds, as implemented in the Large-scale Atomic/Molecular Massively Parallel Simulator (LAMMPS)\cite{plimpton1995fast}.

\begin{figure}
    \centering
    \includegraphics[width=15.0cm] {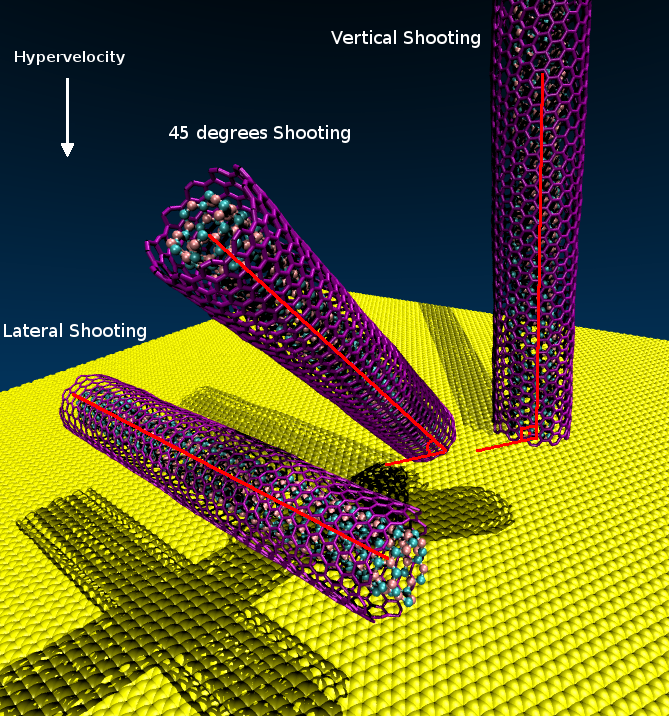}
    \caption{Schematic representation of hypervelocity shooting cases (vertical, lateral and 45 degrees) of the hybrid nanotube against a rigid wall.}
    \label{fig2}
\end{figure}

Initially, all structures, were thermalized at 300K using NVT ensemble and Nose-Hoover thermostat\cite{hoover1985canonical}. After that, they were shot at high-velocity values, varying from 2.4 up to 8.0 km/s, and three impact angles: lateral ($0^o$), $45^o$ and vertical shootings ($90^o$) against a solid wall (Figure \ref{fig2}). This substrate interacts with the nanotube via van der Waals interactions. 

The trajectory of each atom was calculated by integrating Newton's equation ($a_n = \frac{F_n}{m}$), using the numerical method of Velocity-Verlet \cite{ young2014leapfrog, becker2001dynamics}:

\begin{equation}
    r_{n+1} = r_n + v_n\Delta t + \frac{1}{2}a_n\Delta t^2
\label{eq r}
\end{equation}

\begin{equation}
    v_{n+\frac{1}{2}} = v_n + \frac{1}{2}a_n\Delta t
\label{eq frac}
\end{equation} where $r,v,a$ are the position, velocity, and acceleration, respectively. The $\Delta t$ is the timestep, which we used 0.1 fs to all simulations. 
The resolution method consists of:

\begin{enumerate}
    \item Calculate the $r_{n+1}$ using equation \ref{eq r}
    \item Obtain the velocity between $n$ and $n+1$ ($V_{n+\frac{1}{2}}$) with equation \ref{eq frac}
    \item Compute the acceleration $a_{n+1}$, for a time $t + \Delta T$
    \item Finally, obtain the velocity $v_{n+1}$ using:
\end{enumerate}
\begin{equation}
    v_{n+1} = v_n + \frac{1}{2}[a_n + a_{n+1}]\Delta t
\end{equation}

This procedure is repeated till the end of the simulation.

 The stress analysis was also carried out for all simulations, especially due to unzipping possibility that ones related to the lateral impact. This analysis was made using the von Mises stress: 
 \begin{equation}
    \sigma_{vm} = \sqrt{\frac{(\sigma_{11} - \sigma_{22})^2 + (\sigma_{22}  -\sigma_{33})^2 + (\sigma_{11} - \sigma_{33})^2 + 6(\sigma_{12}^2 + \sigma_{23}^2 + \sigma_{31}^2)}{2}}
    \label{eq 2}
\end{equation}

The $\sigma_{ij}$ corresponds to the uniaxial component of stress when $i = j$, and corresponds to the shear stress component when $i\neq j$. Thus have been possible to analyze the stress evaluation along the collision procedure, helping to understand the propagation and the concentration of stress in the hybrid nanotube.

We can compare this stress methodology with the virial stress tensor \cite{zimmerman2004calculation, subramaniyan2008continuum}.
Let $N, m, r, v, f,$ and $V$ be the number of atoms, the mass, the position, the velocity, the force acting on each atom, and the total volume of the system, respectively:

\begin{equation}
    \sigma_{ij} = \frac{\sum_k^N m_k v_{k_i} v_{k_j} + \sum_k^N r_{k_i} f_{k_j}}{V}
    \label{eq 1}
\end{equation} where $\sigma_{ii}$  is the uniaxial component of stress on the ii directions, and it is divided into two terms, the potential, and the kinetic term.

This methodology has been applied with success to study of carbon and BN nanotubes \cite{perim2013dynamical}, carbon nanoscrolls \cite{de2016carbon} under same conditions.

\section{Results and Discussions}

\begin{figure} [h!]
    \centering
    \includegraphics [width = 12.0 cm] {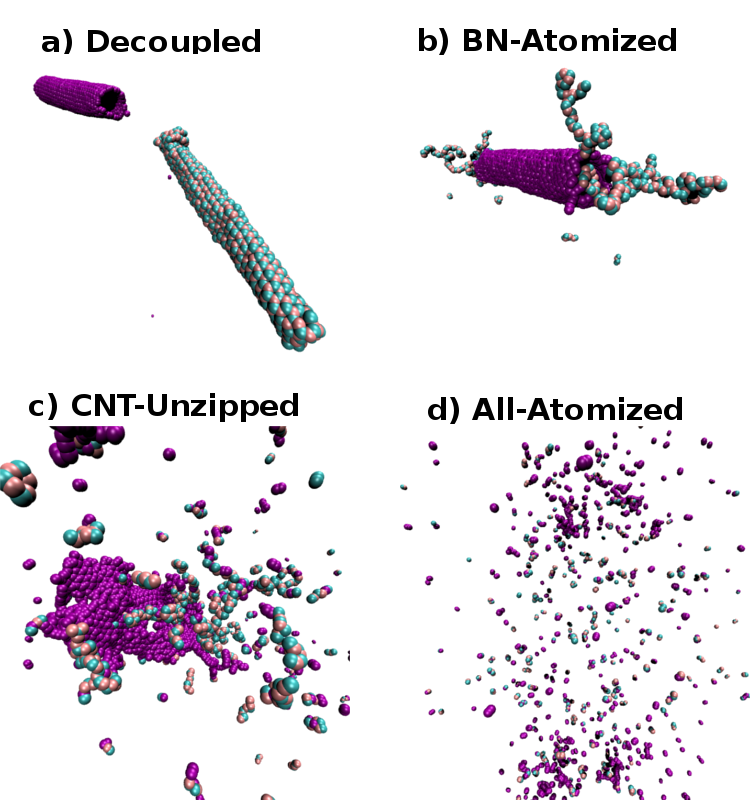}
    \caption{Principal noticeable common results after the impact. Considering all the 30 shooting cases}
    \label{fig3}
\end{figure}

The four representative resulting structures obtained from the high-velocity impact at different angles and velocities are presented in Figure \ref{fig3}. We have named them in accord with the modified nanotube wall. The decoupled structure is that one in which the carbon and BN walls are preserved despite BN launching out CNT. This case is entirely new and has not been reported by other similar works with Double-walled BNNT (DWBNNT) \cite{machado2016structural, perim2013dynamical} and CNT (DWCNT) \cite{dos2012unzipping}.  In the second case, named BN-atomized, the impact of hybrid nanotube atomizes BNNT inside the intact CNT nanotube. This is also a non-reported case in doubled wall BNNT and CNT, and it is because BNNT is brittle than CNT \cite{akdim2003comparative}. Figure \ref{fig3}(c)(d) presented two already reported cases \cite{dos2012unzipping, ozden2014unzipping, de2016carbon} named CNT-Unzipped and All-atomized. Although the similarity \cite{dos2012unzipping, ozden2014unzipping}, CNT-Unzipped from HNT high-velocity impact differentiate by the possibility to obtain only one carbon nanoribbon from a double-walled nanotube. The last case (All-atomized) is the common sense related to the destruction of all walls of the HNT.

\begin{figure}
    \centering
    \includegraphics[width=10.0cm]{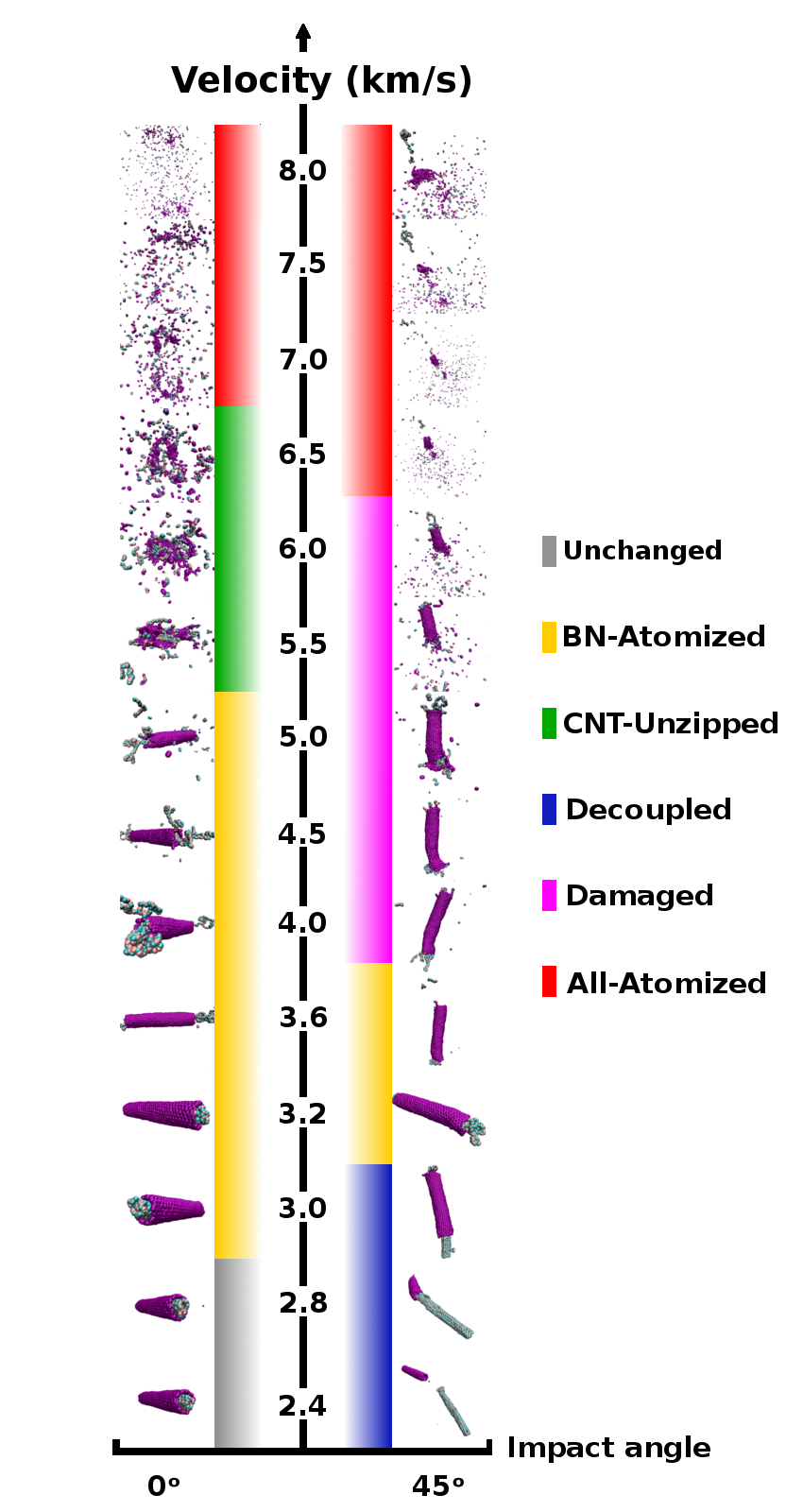}
    \caption{Diagram of the results obtained from all the shooting cases for 0° and 45°. The colors in the columns indicates the classification of the resulting structures, according to the label on the right side of the figure. }
    \label{fig4}
\end{figure}

The conditions to the obtained structures are presented in the Figure \ref{fig4} and show that they are highly dependent on impact angle and velocity. While for 0 degrees, velocities below 3 km/s can not provide HNT structural modifications (unchanged), the asymmetry of 45 impact angle can decouple the walls, resulting in independent CNT and BNT (Decoupled). This phenomenon is intrinsically related to the energy provided in the impact has not been converted in the deformation, as in DWCNT case \cite{machado2016structural}, and sufficient to overcome the BN-C cohesive energy, what is higher than graphene layers \cite{zhou2015van}.The BN atomization can occur in a higher range of velocities in 0 degrees than 45 degrees. It is related to the larger impact area, providing higher stress along the inner nanotube. Figures 1 to 4 in Supplementary material show that atomization can also result in BN linear atomic chains.  Velocities higher than $3.6$ km/s in $45$ degrees impact angle produce only damaged structures. In good agreement with previous works, the CNT-Unzipped can occur just for 0 degree impact angle and in a very specify velocity range: 5.5 to 6.5 km/s. For higher velocities, for all impact angle, the structures are destroyed (all atomized).

\begin{figure}
    \centering
    \includegraphics[width=12.0 cm]{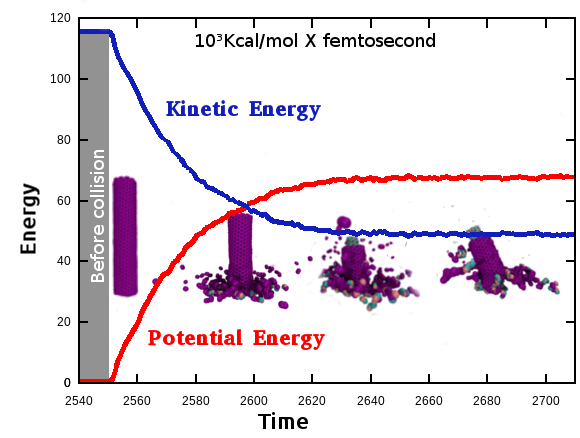}
    \caption{Kinetic and Potential energies during the vertical collision as a function of time.}
    \label{fig5}
\end{figure}

Vertical collisions (90 degrees) always results in partially or totally damaged structures. In Figure \ref{fig5} we show the kinetic and potential energy during the time for a representative case on vertical impact with the respective snapshots. For $6$ km/s, the HNT starts to be destroyed in contact with the substrate, leading to a partially deformed structure. As can be seen, this direction is conducive to the conversion of kinetic to potential energy (deformation). For velocities higher than 7.0 km/s the HNT suffers complete atomization.

\begin{figure}
    \centering
    \includegraphics [width = 12.0cm] {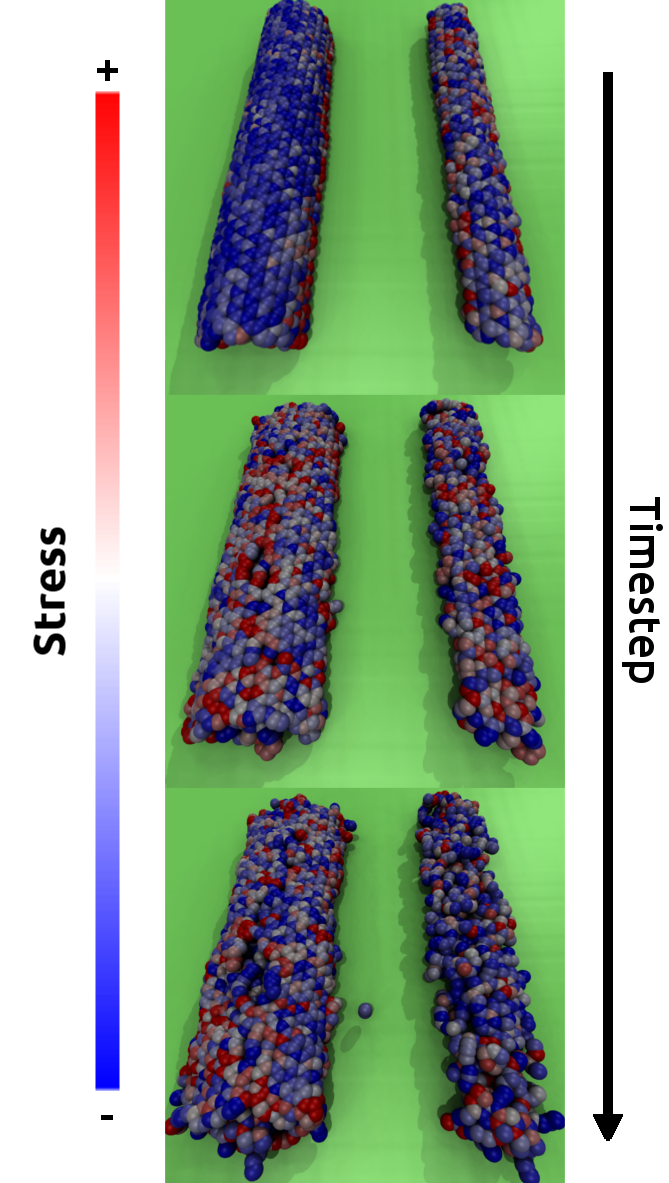}
    \caption{Von Mises Stress concentration along the simulation time in the 6.0km/s lateral shooting case, for the HNT and BNNT, respectively. The atoms are colored in accordance with the scale bar.}
    \label{fig6}
\end{figure}

To gain insights on the HNT unzipping mechanism, we analyze the stress per atom pattern during the impact (Figure \ref{fig6}). As can be seen in Figure \ref{fig6}, the highest stress concentration in CNT (HNT outlayer) differs remarkably from that one in DWCNT \cite{ozden2014unzipping}. For DWCNT, the stress concentration occurs in edges what lead to the formation of a bilayer graphene. In the other hand, for HNT, the stress concentration occurs in the center, similiarly to the SWCNT \cite{ozden2014unzipping}, driving to a graphene membrane formation. Similar observations are presented in the BNNT (HNT innerlayer) but leading to the amorphization.

\begin{figure}
    \centering
    \includegraphics[width = 10.0cm] {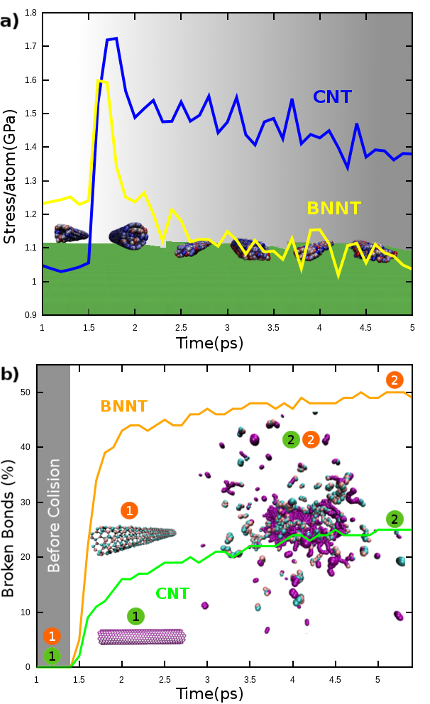}
    \caption{(a) The stress of each atom as a function of time with representative snapshots of the 6.0 km/s 0° impact. (b) Percentage of broken bonds per nanotube along HN the collision trajectory.}
    \label{fig7}
\end{figure}

The Figure \ref{fig7} presents the average stress as a function of time (Figure \ref{fig7} a)). It is possible to observe that the maximum stress in BNNT is lower than in CNT. The fast BN breaking bonds (Figure \ref{fig7} b)) and new structures formations lead to a fast stress decay, lower than the initial one. Those results suggest the formation of stabler BN structures. In CNT, the percentage of breaking bonds (Figure \ref{fig7} b)) is lower than in BNNT which difficulties the stress release and is an indicative of the honeycomb graphene structure maintaining.

\section{Conclusion}

In summary, our MD simulations revealed that controlling collision velocities and the impact angles it is possible to obtain a greater variety of structures than in pure MWCNT and BNNT at same conditions \cite{machado2016structural, perim2013dynamical,dos2012unzipping,ozden2014unzipping}. Nanoribbons from CNT unzipping is also obtained for HNT at high velocity, but the mechanism contrasts with the pure carbon double walled \cite{dos2012unzipping} nanotubes once the stress propagation occurs from bottom to the top opening the CNT at the middle, due to BN atomization inside the nanotube. This process is similar to doubled walled BN nanotubes unzipping.
Impact occurring in angles bigger than 0 and velocities around 2.4 to 3.0 km/s, eject intact BNNT from CNT. Bigger velocities (3.6 to 4.5 km/s) at lateral impact can produce BN linear chain in the CNT.

\section{Acknowledgements}
The authors thank the Brazilian agency CNPq for financial support. The
authors also acknowledge the Federal University of ABC for computational support.







\section*{References}
\bibliographystyle{model}
\bibliography{references.bib}







\end{document}